\documentclass[preprint,review,11pt]{elsarticle}

\usepackage{graphicx}
\usepackage{amsmath}
\usepackage{amssymb}

\usepackage{lineno}

\usepackage{algpseudocode}
\usepackage{algorithm}

\usepackage{caption}
\usepackage{subcaption}
\usepackage{fixltx2e}

\usepackage{xcolor}
\usepackage{multirow}

\usepackage{cellspace}
\newlength{\bibitemsep}
\newlength{\bibparskip}
\let\oldthebibliography\thebibliography
\renewcommand\thebibliography[1]{%
  \oldthebibliography{#1}%
  \setlength{\parskip}{\bibitemsep}%
  \setlength{\itemsep}{\bibparskip}%
}

\usepackage{etoolbox}
\makeatletter
\expandafter\patchcmd\csname\string\algorithmic\endcsname{\itemsep\z@}{\itemsep=-0.7ex plus2pt}{}{}
\makeatother

\journal{Pervasive and Mobile Computing Journal}

\begin{document}

\begin{frontmatter}

\title{Cooperative Monitoring and Dissemination of Urban Events Supported by Dynamic Clustering of Vehicles}

\author[mymainaddress]{Everaldo Andrade}

\author[mymainaddress]{Kevin Veloso}

\author[mymainaddress]{Nath\'alia Vasconcelos}

\author[mysecondaryaddress]{Aldri Santos}

\author[mymainaddress]{Fernando Matos\corref{mycorrespondingauthor}}
\cortext[mycorrespondingauthor]{Corresponding author}
\ead{fernando@ci.ufpb.br}

\address[mymainaddress]{Centre of Informatics, Federal University of Para\'iba, Jo\~ao Pessoa, Brazil}
\address[mysecondaryaddress]{Department of Informatics, Federal University of Paran\'a, Curitiba, Brazil}

\begin{abstract}

Critical urban events take places at a random way and they need to be dealt with by public authorities quickly to maintain the proper operation of cities. The main challenges for an efficient handling of an event fall precisely in its random nature, and in the speed and accuracy of the notification of its manifestation to the authority. The pervasiveness of vehicles in urban environments, and their communication and monitoring capabilities, allow to employ VANETs to support the dissemination and handling of such events. Thus, this work proposes MINUET, a dynamic system to support the monitoring and dissemination of urban events, which operates in an ad hoc vehicular network. It uses a cooperative-based strategy where vehicles are able to dynamically coordinate the monitoring and dissemination of events in real time by the means of clusters. Results obtained by NS3 show the performance of MINUET regarding the real-time monitoring and dissemination of events detected in urban areas.

\end{abstract}


\end{frontmatter}

\section{Introduction}
\label{sec:intro}

In recent years, urban centers have experienced unprecedented population growth. This growth complicates several everyday aspects of the city, such as urban mobility~\cite{olaverri_fugure_2018}, thus making it more difficult for public authorities to manage public services. One of the problems faced by city managers is in handling critical urban events by the means of detection, analysis and monitoring of such events~\cite{memos_efficient_2018}. These events can be of several types, such as fires, accidents, assaults, cars on the run, wanted persons or vehicles, blockades, public protests, among many others. The prompt and efficient handling of these events, through the collaboration among agents distributed in the environment is essential to guarantee the safety and assistance of the individuals that are directly or indirectly involved~\cite{monajemi_information_2018}. In addition, a quick response ensures a smooth operation of the city activities. For instance, when detecting a blockade on an avenue, the public mobility entity must act quickly to clear the way so it does not create long and time-consuming congestion in other parts of the city.

The Infrastructure and Communication Technologies (ICTs) of the Smart Cities can play a crucial role regarding the handling of the urban events. For instance, by enhancing the intercommunication between the public infrastructure and the authorities it is possible to deliver smarter services, such as monitoring and alert services, to improve the quality of life of citizens \cite{alsamhi_survey_2019, fernandez_studying_2017}. Moreover, with the evolution of 
Crowdsensing and IoT, the integration and cooperation of different computational systems may allow the advent of new management strategies of emergence response and monitoring in urban zones. Hence, it is possible to idealize resilient infrastructures to guarantee the city operation continuity even in face of catastrophic events~\cite{timashev_resilient_2017,furutai_resilience_2018}. Usually, those strategies employ devices that can collect data (e.g. smartphones, cameras and vehicles) to support other systems in their the decision making processes regarding the treatment of specific~events.

Intelligent Transport Systems (ITS) contribute to the automation of movement operations within Intelligent Cities. In ITSs, vehicles are part of a Vehicular Ad Hoc Network (VANET), acting as mobile crowd sensors \cite{kong_mobile_2020, ren_minimum_2018} in urban sensing scenarios to assist the development of smart applications \cite{ali_dissemination_2014}. They intercommunicate with other vehicles (Vehicle-to-Vehicle - V2V) and the city's network infrastructure (Vehicle-to-Infrastructure - V2I) to exchange and disseminate contextual information \cite{bazzi2018}, which is paramount for a variety of safety and entertainment related goals \cite{fernandez_studying_2017, talib_2019}. In addition, due to their pervasiveness in urban centers \cite{bettini_privacy_2015}, vehicles seem to be an obvious option for the collection of multimedia data of events~\cite{wang_data_2018, quadros_beacon-less_2015, quadros_qoe-driven_2016} and to support cooperation and coordination in monitoring processes~\cite{castro_survey_2017}, since they do not suffer from the limited resources that other ad hoc networks have, such as low processing power, low storage capacity  and battery dependency~\cite{khakpour_using_2017}. However, VANETs impose their own challenges, such as high mobility~\cite{silva_cognitive_2016}, frequent handover \cite{alam2019}, variation of traffic density, high scalability and network topological variation~\cite{macedo_experimental_2014}. In this context, recent studies have demonstrated a variety of strategies to manage urban events. However, such strategies usually handle a specific event only, such flooding~\cite{Mousa2016}, vehicular accidents~\cite{chou_adaptive_2017}, traffic condition warnings~\cite{Abhishek2016}. Other works focus on the dissemination of surveillance~\cite{huang_evac-av:_2017} or collision~\cite{shi_centralized_2017} data, though, by employing a centralized approach.

This work presents a dynamic system based on cooperative clustering of vehicles called MINUET (\textbf{M}onitorINg and Dissemination of \textbf{U}rban \textbf{E}ven\textbf{T}s) that carries out on a hybrid vehicular network by employing V2V and V2I communication. MINUET employs a decentralized approach to support the coordination and control of dynamic monitoring and dissemination of multiple urban events (e.g. accidents, disasters and robberies) in order to assist public authorities in the identification of emergency events. In the proposed model, vehicles close to an emergency event self-coordinate to monitor it as long as possible in order to inform a public entity, so it can take the most effective decision to resolve it. In addition, vehicles maintain collaborative monitoring and cooperative communication, where multiple user resources (vehicles) transmit data from a single source (event) and there is a combination at the destination (competent entity) of data transmitted by cooperative users~\cite{hong_cooperative_2007}. Once in the destination, this data can be interpreted, generating information to assist the entity in the decision making process regarding the resolution of the event. Thus, the purpose of this~coo\-perative communication is to allow the dissemination of event information to the competent entities to be performed in a more efficient fashion. MINUET~was evaluated through simulations, by employing the clustering techniques proposed in~\cite{tal_novel_2016} and \cite{khakpour_using_2017} to measure the efficacy and efficiency of the solution. The results showed that MINUET can detect, monitor~and disseminate event data by the means of clustering, thus supporting the treatment of these events.

The remainder of this article is structured as follows: Section~\ref{sec:related} presents the related work. Section~\ref{sec:proposal} describes the MINUET system and its operation, while Section~\ref{sec:eval} details the evaluation applied in the analysis of the MINUET performance. Lastly, Section ~\ref{sec:concl} concludes the work with our~considerations.

\vspace{-3mm}
\section{Related Work}
\label{sec:related}
The pervasiveness of vehicles in the cities was one of the causes that led to the increasing of studies in VANETs. Several solutions in the literature use VANETs to support the daily life of cities when dealing with specific urban aspects, targeting safety and non-safety goals, such as accident prevention and traffic efficiency, respectively \cite{alam2019}. Some solutions for smart cities use clustering to support the tracking of target objects and/or the dissemination of information. Table \ref{tab:related} presents a comparison of the related works taking into account the characteristics we believe are important 
in the management of urban events. This table assists in perceiving the MINUET contributions by showing the limitations of each work.

A control scheme that uses vision-area-based clustering algorithm together with adaptive video streams techniques has been proposed in~\cite{huang_evac-av:_2017}. This proposal aims to assist in surveillance of the traffic lanes, in which vehicle groups propagate visual information in real time. The solution seeks to choose the most appropriate group leader, who captures images of the route and passes it on to the other members. However, only the leader captures the information, thus not allowing collaboration during monitoring. In~\cite{siddiqua_icafe_2019} the authors propose a content centric protocol for congestion avoidance and fast emergence (iCAFE). When a vehicle is affected by a critical condition (e.g. colision), it sends an emergency message to RSUs to initiate the process of alerting a nearby ambulance. Despite the good simulated results, iCAFE does not ensure a collaborative monitoring, since only the affected vehicle triggers the emergence message. In case the OBU is damaged during the accident, the rescue process is not initiated.

The authors in \cite{shi_centralized_2017} propose a centralized clustering solution based on a hybrid network architecture. By using IEEE 802.11p and LTE interfaces, vehicles choose the communication model and disseminate alert messages to other network groups through a central entity. Such messages alert drivers to intersection collisions and traffic jams. However, the central entity is the only responsible for cluster formation and maintenance, thus limiting the self-coordination of the vehicles. In addition, only the cluster leader communicate with the structured network, which can cause delays in the dissemination of information. In~\cite{yedder_reactive_2017} the authors propose a information dissemination reactive system to aid in the dispatch of emergency vehicles. From information collected by vehicles in VANETs, it is possible to assume that a road is congested, which allows the route of the emergency vehicle to be changed in real time. However, simulations are only performed considering mobility and traffic issues, without regard to network infrastructure issues (e.g. delay and collisions) that could influence the solution. Besides, it employs a centralized agent to coordinate the actions.

In~\cite{wang_optimizing_2018} the authors propose a framework for content dissemination in heterogeneous access networks using a group-based traffic management strategy. The work uses a crowdsensing-based system to collect event information in a collaborative way to deliver it to a central management. Group members send messages about an event to the leader of the group. The leader analyzes the data in order to have more accurate information about the events and then decides whether to send it immediately to the central management or not. The work presented in~\cite{benkerdagh_cluster_2019} proposes a solution for event alert dissemination based on optimization and clustering techniques. Instead using full textual messages, it employs codes to decrease the number of exchanged packets. In addition, it ensures the clusters stability by using a fitness function to choose the vehicles that are part of the clusters. One problem is that only vehicles that move towards the event can participate in the cluster, thus limiting the cooperative communication.

A decentralized strategy for the acquisition of phenomena in vehicular networks is proposed in ~\cite{gorrieri_clustering_2016}. In a initial phase (downlink phase), a remote sink node, which may reside in the cloud, triggers messages in the network for group formation. During the formation, the leaders of each group are already defined. In a second phase (uplink phase), data collected by the vehicles of each group, referring to the observed phenomenon, are sent to their respective leaders and transferred to the sink node. However, this solution employs a proactive strategy that can waste network resources in the absence of  phenomenon. 

\begin{table}[h]
\renewcommand{\arraystretch}{0.8}
\centering
\footnotesize
\caption{Comparison of related works about the characteristics required}

\label{tab:related}
\begin{tabular}{@{\extracolsep{2pt}}lp{2.5cm}p{2.4cm}p{2.3cm}p{2cm}}
\hline
\textbf{Ref.} & \textbf{Cooperative communication} & \textbf{Collaborative monitoring} & \textbf{Self-coordination} & \textbf{Resource saving}\\ \hline
\cite{huang_evac-av:_2017} & Yes & {\it No} & Yes & Yes\\
\cite{siddiqua_icafe_2019} & Yes & {\it No} & Yes & Yes\\
\cite{shi_centralized_2017} & {\it No} & Yes & No & Yes\\
\cite{yedder_reactive_2017} & Yes & Yes & {\it Limited} & Yes\\
\cite{wang_optimizing_2018} & {\it No} & Yes & Yes & Yes\\
\cite{benkerdagh_cluster_2019} & {\it Limited} & Yes & Yes & Yes\\
\cite{gorrieri_clustering_2016} & Yes & Yes & Yes & {\it No} \\
\hline
\end{tabular}
\vspace{-4mm}
\end{table}

\vspace{-3mm}
\section{Detection and Dissemination of Urban Critical Events System}
\label{sec:proposal}

This section introduces the MINUET system to support the cooperative monitoring of critical urban events through on-demand vehicles clustering. It operates on hybrid vehicular networks (V2V and V2I communication) to support the coordination and control of critical event monitoring and dissemination of collected data to an external entity that identifies and handles urban events. In the external entity, the data is analyzed by target applications in order to assist the entity in making decisions to resolve the events. Initially, we describe 
the urban environment~model of MINUET. Next, we detail the MINUET architecture, and the components interactions, and illustrate their operation. 

\subsection{Urban Environment Model}
\label{sec:urban}

\begin{figure}[h]
	\centering
	\includegraphics[scale=0.35]{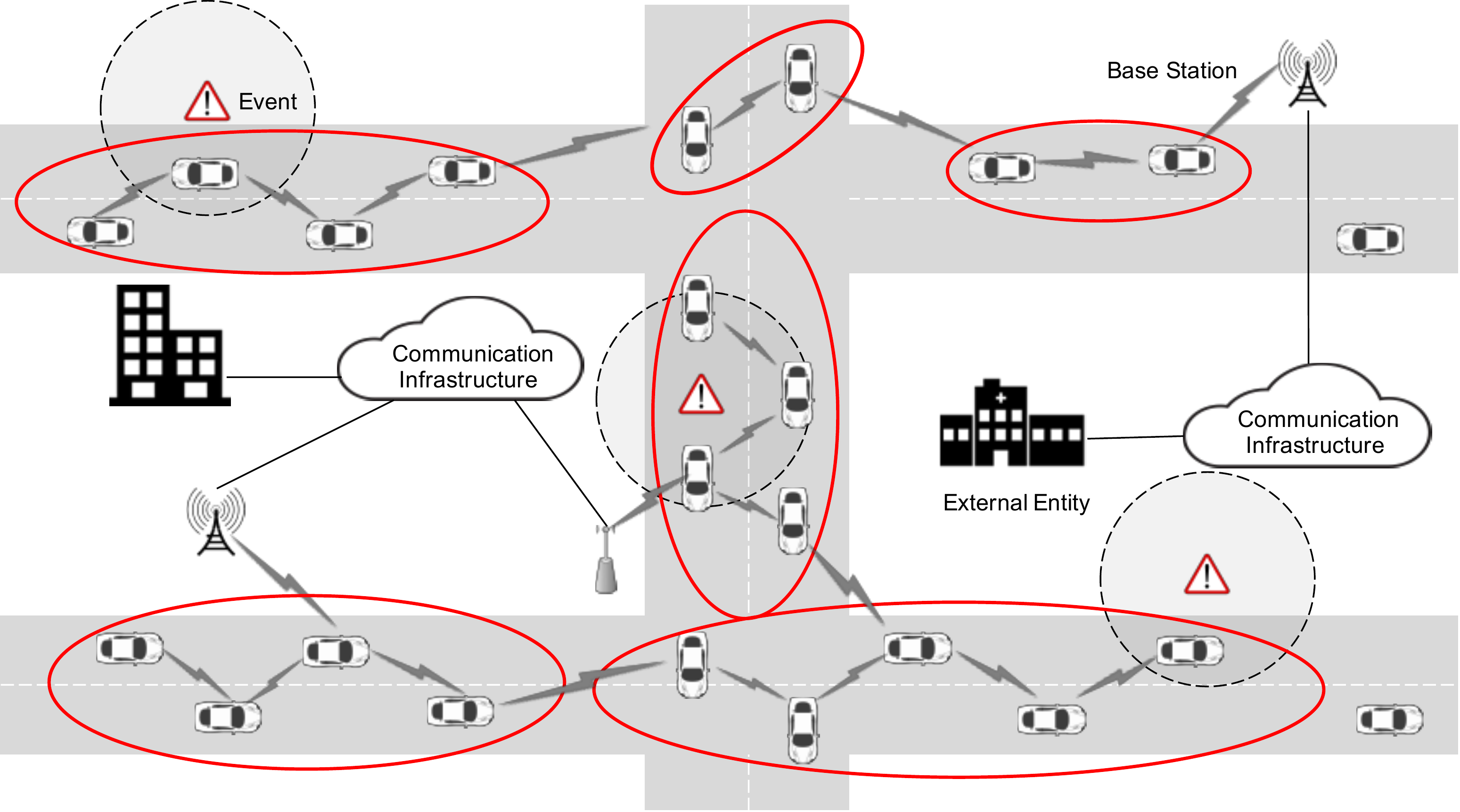}
	\caption{Urban environment model}
    \vspace{-5mm}
	\label{fig:urban_model}
\end{figure}

The urban environment where MINUET operates consists of an unstructured geographic space formed by structures and road networks where vehicles move continuously obeying traffic laws, as illustrated in Figure~\ref{fig:urban_model}. All vehicles are able to communicate with each other (V2V) and with the urban infrastructure (V2I) through Base Stations (BS), thus creating hybrid networks. As it happens in real urban areas, critical urban events of distinct natures may arise randomly in time and space. A critical urban event is any event that may impact on the day-to-day of the city and its citizens, such as fires, accidents, crimes, obstructions in the streets, among others. These events must be treated by an External Entity (EE), representing a legal authority, which communicates through the BSs with the~VANET.

The urban environment model is composed of a set $V$ of $n$ vehicles (nodes) denoted by $\{v_{1}, v_{2},...,v_{n}\}$ and a set $E$ of $m$ events denoted by $\{e_1, e_2,...,e_m\}$, where any vehicle $v_i \in V$ can detect a event $e_j \in E$ if $e_j$ is within the range of $v_i$. This detection is possible due to georeferencing devices and/or embedded sensors and $v_i$ can communicate with other vehicles or with the urban infrastructure by the means of its communication devices. Every event has a lifetime that begins at the moment the event occurs until it vanishes or it is handled by an acknowledged authority. An event has a position in space, which can vary over the time, thus typifying a fixed or mobile event. Once detected, the event must be monitored with the support of a group of vehicles. Therefore, $E' \subset E$ is the set of events that were detected and for each $ ev \in E'$, it is defined a set of monitoring vehicles $Gm(ev) = \{vm\}$, where $ vm \subset V$. The urban model then has a set $G$ of $r$ monitoring groups denoted by $\{ Gm_1, Gm_2,...,Gm_r\}$. It is worth stressing that the size of a monitoring group associated to an event is dynamic, since it can change over time due to the mobility of the vehicles or the event itself. In addition, vehicles can join and leave groups, depending on the vehicles' density on the roads and their directions.

In Figure~\ref{fig:urban_model}, the red ellipses mean groups of monitoring vehicles responsible for monitoring and disseminating information on occurring events at a given time. The event mobility must be taken into account since the groups must remain close to their respective events. Therefore, a group responsible for monitoring a mobile event should follow it as long as possible, while a group responsible for monitoring a fixed event is likely to move away from the respective event. Thus, due to the high dynamics of the vehicular network and events' mobility, events can be detected at intermittent intervals, which may result in interruption on monitoring.

Finally, in order to better treat the event, the monitored data must be delivered to the EE in a maximum time limit so as to be usable by their respective target applications, and thus be useful to the EE when making decisions about the event resolution. Obviously, this time limit depends directly on the nature of the event. For example, the time limit for delivering the monitored data of a road blockage event can be higher than that associated to a car on the run event. The maximum delivery time $ MDT(ev) $ is then defined as the maximum time interval that the monitored data of $ ev \in E'$ must be delivered to the destination.

\subsection{MINUET} 
\label{sec:prop}
\begin{figure}[h]
	\centering
	\includegraphics[scale=0.5]{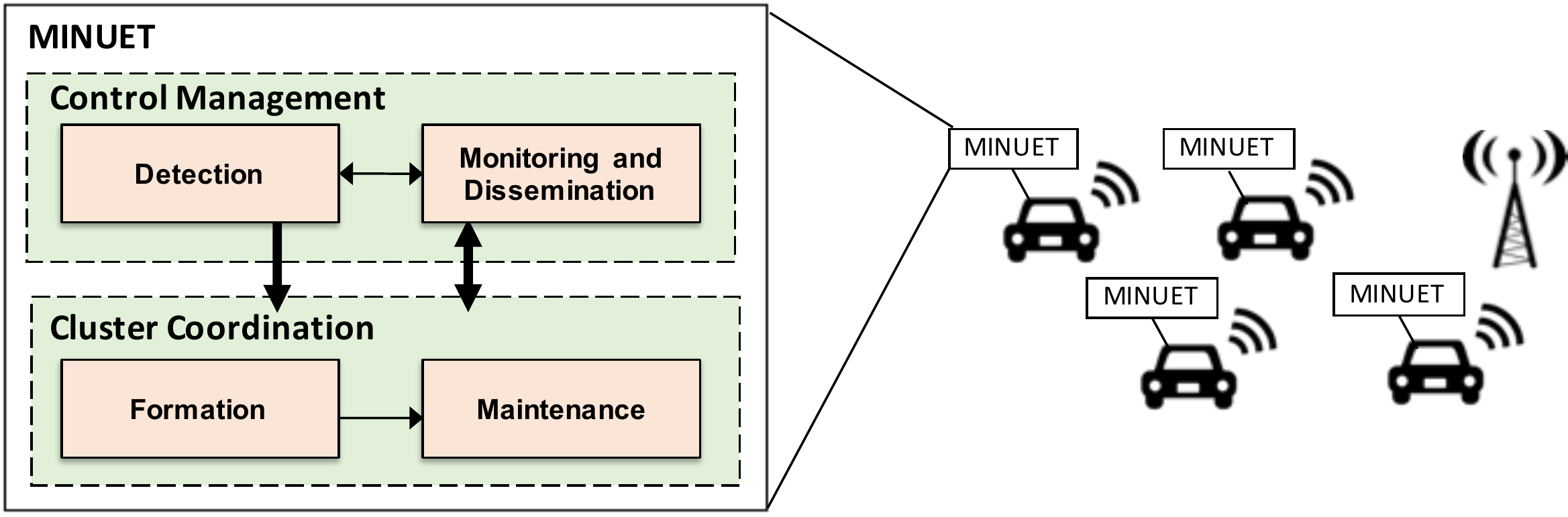}
	\caption{The MINUET architecture}
    \vspace{-5mm}
	\label{fig:architecture}
\end{figure}

The MINUET (\textbf{M}onitorINg and Dissemination of \textbf{U}rban \textbf{E}ven\textbf{T}s) system is proposed to provide cooperative and dynamic event monitoring supported by groups of vehicles. Its architecture consists of two main modules, called \textbf{Cluster Coordination} and \textbf{Control Management}, as shown in Figure~\ref{fig:architecture}. The \textbf{Cluster Coordination Module} configures the formation and maintenance of vehicle clusters that will monitor the event. The \textbf{Control Management Module} is responsible for coordinating the detection, announcement, and monitoring of events and the dissemination of information from these events to the EEs.

\textbf{The Cluster Coordination Module} aims to create and maintain clusters of vehicles and consists of two components: \textit{Formation} and \textit{Maintenance}. The \textit{Formation} component is responsible for discovering neighbors, selecting the cluster leader, defining the roles of each node in the cluster and establishing the communication of the nodes with their respective leaders. Initially, a node should periodically announce its existence to its neighbors while simultaneously obtaining similar information from them in order to build the network topology. Based on information from neighbors, nodes may or may not be part of a cluster. Those which are chosen to be part of the cluster are called member nodes. In addition, a node is selected as a leader to perform cluster maintenance operations.

Once the cluster is created, the \textit{Maintenance} component coordinates the processes that occur when nodes leave or join the cluster. When considering the high node speed, event mobility, network topological variation, conductor behavior and channel interference, the number of nodes in the clusters can vary considerably over time. Thus, it is necessary to create mechanisms to deal with this high network dynamicity without compromising its performance. This component also deals with overlapping clusters, that is when two or more clusters share the same region at the same time. In such cases, it may be necessary to merge these overlapping clusters. It is worth stressing that different clustering techniques can be used, thus each technique defines the criteria for cluster formation and maintenance.

\textbf{The Control Management Module} addresses the detection, the event monitoring and the dissemination of the information to the EE. It is composed of two components, where the first one deals with the detection and announcement of the event, while the second component handles the monitoring and dissemination of the event information. To accomplish that, it is also necessary to manage and control the monitoring in the cluster, apart from the management of the cluster itself. Vehicles must dynamically self-coordinate to collaborate in the treatment of multiple events on different traffic scenarios in a metropolitan region. Thus, in addition to the roles defined by the \textit {Cluster Coordination} module, this module also defines three other roles, which are: \textit{(i) Monitor}: if the event is in the field of view of the vehicle, than the vehicle can monitor the event. \textit{(ii) Transmitter}: vehicle that disseminates the information within the cluster only, in order to reach the EE. \textit{(iii) Gateway}: vehicle that delivers the information of the monitored event to the BS and, consequently, to the EE. The node \textit{gateway} also establishes communication between clusters. Vehicles can play more than one role simultaneously. For instance, a vehicle can be both a \textit{monitor}, since it is close to the event, and a \textit{gateway}, because it is within range of the BS.

Initially, the \textit{Detection} component handles the analysis and detection of an event by a vehicle. A vehicle \textit{vd} on the road, can detect the occurrence of an event \textit{ev} at a given time. After the detection, \textit{vd} collects context data of \textit{ev}, which are \textit{time of detection}, \textit{position}, \textit{velocity} and \textit{direction}. For the sake of simplicity, it is assumed that the analysis and detection of the event are performed through image processing and analysis techniques, which run on cameras embedded on vehicles. In addition, it is also assumed that during the analysis of the event, \textit{vd} can estimate the $MDT(ev)$.

Once the context information is collected, announcement messages are disseminated through the network by employing a flooding strategy. However, not all vehicles need to be aware of the event occurrence, and after some time, the event announcement is no longer useful. Therefore, it is recommended to limit the dissemination of the event information to reduce network overhead and possible packet loss and delay. Assuming that a vehicle $vr$ receives the announcement message at time $tr$ and that the event was detected at time $td$, then $vr$ forwards the announcement message if the remaining time $vr$ has to disseminate the message has not yet exceeded the maximum delivery time of the event ($tr - td \leq MDT(ev)$). Otherwise, the message is discarded. This time limit calculated by each vehicle receiving the announcement message delimits an area called Announcement Zone (AZ), which bounds the vehicles that can be members of the cluster that collaborate to monitor the event.

\begin{figure}[h]
	\centering
	\includegraphics[scale=0.25]{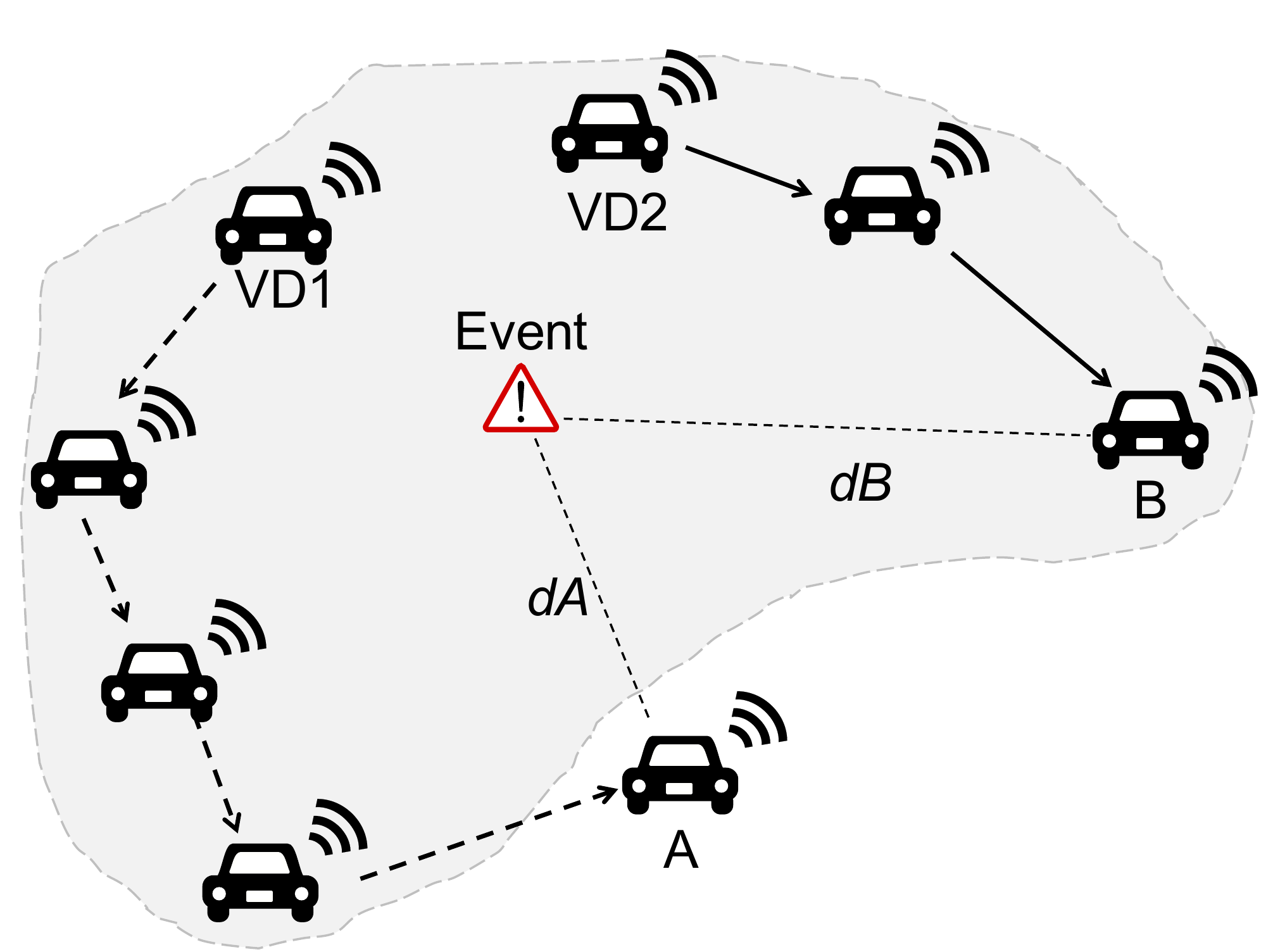}
	\caption{Announcement zone}
\vspace{-5mm}
	\label{fig:zr-za}
\end{figure}

Figure \ref{fig:zr-za} illustrates a detected event and the AZ generated based on this event, as well as the vehicles that are part of the AZ. In this example, vehicles \textit{VD1} and \textit{VD2} detect the event and the announcement messages follow the routes designated by the arrows. It is easy to see that, although vehicle \textit{A} is geographically closer to the event than vehicle \textit{B} ($dA <dB$), \textit{B} is part of the AZ, while \textit{A} is not. This situation occurs when the announcement message takes more time to reach \textit{A} than \textit{B}, which can cause the $MDT$ to be exceeded in \textit{A}. In Figure \ref{fig:zr-za}, the route that the message follows to reach \textit{A} has more hops than the route to \textit{B}, thus increasing the delay and causing the message to be discarded in \textit{A}. All other vehicles continue to broadcast the message. After a vehicle receives an announcement message and confirms that it is in the AZ, it then checks to see if it can be a member of the cluster that collaborate to monitor the event. This step is performed by the \textit{clustering} module, according to the clustering technique designed and used to support the dissemination of the information.

Finally, the \textit {Monitoring and Dissemination} component deals with the monitoring of the event and the dissemination of its data through the cluster so it can be delivered to a BS. To accomplish that, this component employs a self-scheduling scheme, where vehicles alternate between different roles to support the cooperative monitoring. When vehicle \textit{vd} detects the event \textit{ev}, it disseminates announcement messages, and at the same time, initiates the \textit{ev} monitoring by collecting its data and disseminating it by the means of monitoring messages using flooding. While \textit{ev} is within \textit{vd's} range, \textit{vd} plays the role of a \textit{monitor}. If \textit{vd} is also within range of an BS, the collected data is delivered, thus making \textit{vd} to also play the role of a \textit{gateway}. Otherwise, \textit{vd} only disseminates the messages to its neighbors. Each vehicle that receives the monitoring message checks if it is a member of the cluster. If so, it also disseminates the messages, playing the role of a \textit{transmitter}. If the vehicle is within range of an BS, then it plays the role of a \textit{gateway}, thus delivering the monitoring messages.

\begin{algorithm}
 \footnotesize
 \caption{Event detection}
 \label{alg:detection}
 \begin{flushleft}
\vspace{-3mm}
\textbf{Input:} set $V$ of vehicles, detected event $ev$\\
\vspace{-3mm}
 \end{flushleft}
 \begin{algorithmic}[1]
 \For{each vehicle $v_i$ in $V$}
 \If{$ev$ is in $v_i$'s range} \Comment{ $v_i$ is \textit{monitor}}
 \State Start clustering. Add $v_i$ to $Gm(ev) $ \Comment{Request to the Clustering module}
 \State $v_i$ creates and disseminates $msg.Ann(ev)$ and $msg.Mon(ev)$
 \If{$v_i$ is within a $BS's$ range}
 \State Deliver $msg.Mon(ev)$ to the $BS$ \Comment{ $v_i$ is \textit{Gateway}}
 \EndIf
 \EndIf
 \EndFor
 \end{algorithmic}
 \end{algorithm}
 
Algorithm~\ref{alg:detection} describes the process that occurs in vehicles that detect the event $ev$. Once $ev$ is detected, vehicle $v_i$ plays the role of \textit{monitor} and initiates the cluster formation, which is performed by the Clustering module. Then, $v_i$ creates and disseminates announcement and monitoring messages. A announcement message (\textit{msg.Ann(ev)}) contains the event context information needed to maintain the cluster, such as event position, detection time, and event type. A monitoring message (\textit{msg.Mon(ev)}) contains the monitored event data. In addition, if $v_i$ is within a BS's range, $v_i$ also plays the role of \textit{gateway}, thus delivering the \textit{msg.Mon(ev)} to that BS.

\begin{algorithm}
\footnotesize
\caption{Event dissemination}
\label{alg:dissemination}
\begin{flushleft}
\vspace{-3mm}
\textbf{Input:} set $V$ of vehicles, set $Gm(ev)$ of vehicles that monitor $ev$, announcement or monitoring message $msg$\\
\vspace{-3mm}
\end{flushleft}
\begin{algorithmic}[1]
\For{each vehicle $v_i$ in $V$ that receives $msg$}
\State Calculate $ZA(ev)$
\If{$v_i$ it is within $ZA(ev)$} \Comment{If it is not within $ZA(ev)$, discard $msg$}
\If{$msg == msgAnn(ev)$} \Comment{$v_i$ received announcement message}
\State Disseminates $msg.Ann(ev)$
\State Checks if $v_i$ can join $Gm(ev)$ \Comment{Request to the Clustering module}
\Else \Comment{$v_i$ received monitoring message}
\If{$v_i \in Gm(ev)$} \Comment{If it is not member of $Gm(ev)$, discard $msg$}
\State Disseminates $msg.Mon(ev)$ \Comment{ $v_i$ is \textit{Transmitter}}
\If{$v_i$ is within a $BS's$ range}
\State Deliver $msg.Mon(ev)$ to the $BS$ \Comment{ $v_i$ is \textit{Gateway}}
\EndIf
\EndIf
\EndIf
\EndIf
\EndFor
\end{algorithmic}
\end{algorithm}
\vspace{-5mm}

Algorithm \ref{alg:dissemination} determines if the vehicles that did not detect the event and that received announcement or monitoring messages will join the cluster to monitor the event and disseminate the messages. Every time a vehicle $v_i$ receives a message, it checks if it is within the $AZ$ of $ev$. If not, $v_i$ discards the message. Otherwise, if $v_i$ receives an announcement message, it disseminates the message and checks (via the Clustering module) whether it can join $Gm(ev)$. If $v_i$ receives a monitoring message, then it only disseminates the message if it is already member of $Gm(ev)$, thus turning $v_i$ into \textit{transmitter}. In addition, if $v_i$ is within a $BS's$ range, it also plays the role of \textit{gateway} and delivers the message to the $BS$. Finally, if $v_i$ is not member of $Gm(ev)$, then it discards the message.

\begin{figure}[h]
	\centering
	\includegraphics[scale=0.4]{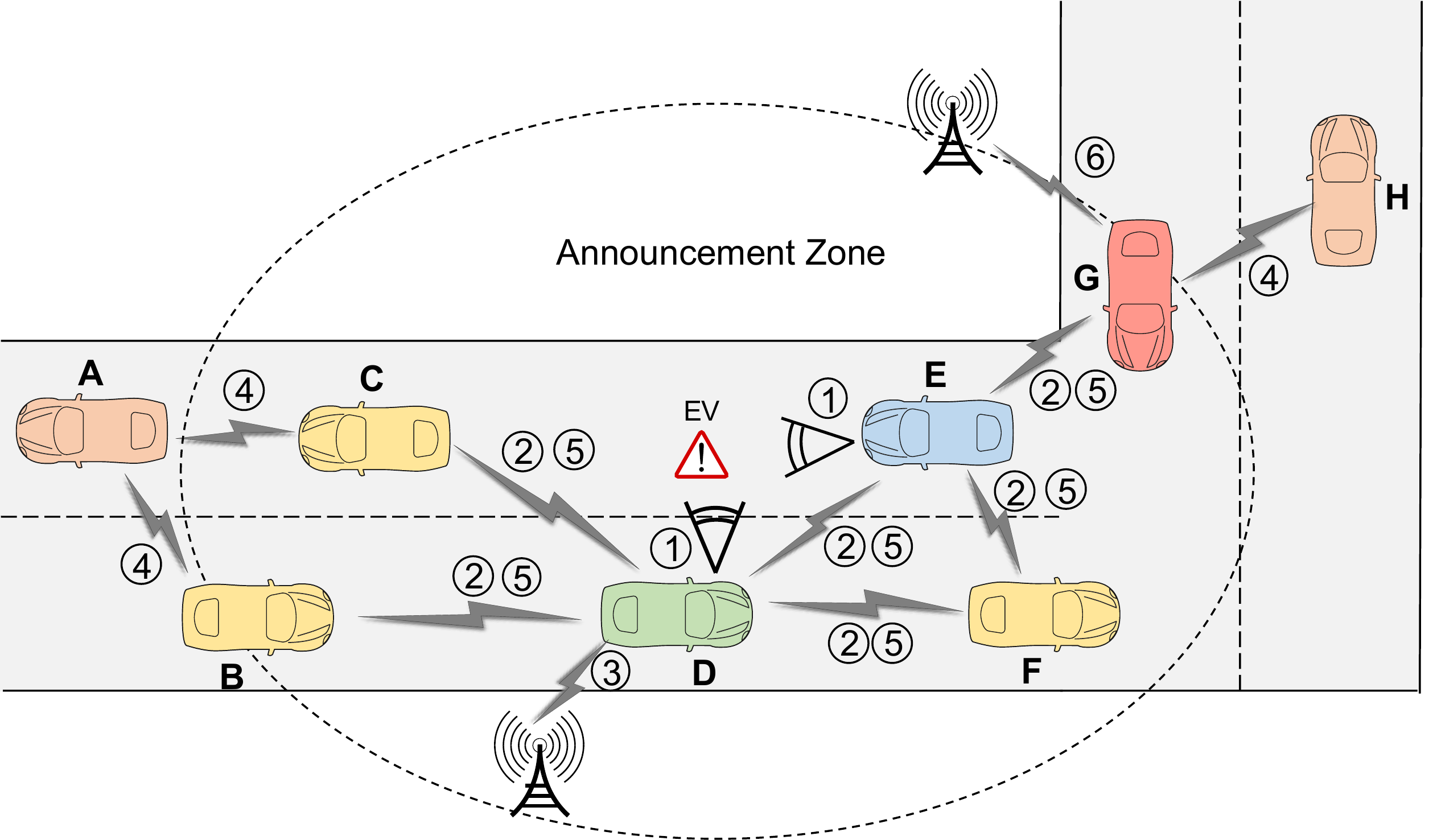}
	\caption{Example of the MINUET execution}
	\label{fig:exemplo}
\vspace*{-5mm}
\end{figure}

Figure~\ref{fig:exemplo} illustrates the MINUET operation scenario containing the set $V=\{A, B, C, D, E, F, G, H\}$ and the set $E=\{EV\}$. Vehicles \textit{D} and \textit{E} can detect the event \textit{EV}, thus creating the set $E'=\{EV\}$. \textit{D} and \textit{E} then collect the context data of \textit{EV} (1), thus playing the role of \textit{monitors}. Immediately, they initiate the dissemination of announcement messages (2). Because \textit{D} is within a $BS's$ range, it also delivers the monitored data (3), thus also becoming a \textit{gateway}. As announcement messages are being disseminated, the vehicles verify if they are within the AZ and if they are members of the cluster created to perform the cooperative monitoring. Upon receiving the announcement message (4), vehicles \textit{A} and \textit{H} verify that they are not within the AZ, thus not participating in the monitoring of \textit{EV}. For the sake of simplicity, it is assumed that all vehicles that are in the AZ are members of the cluster. Thus, the cluster $Gm(EV)=\{B, C, D, E, F, G\}$ is created. As the group is being created, monitoring messages are also disseminated through its members (5). Vehicles \textit{B}, \textit{C} and \textit{F} then play the role of \textit{transmitters} while vehicle \textit{G} plays the gateway role, since it can deliver the monitoring message to a BS (6).

\vspace{-3mm}
\section{Evaluation}
This section presents a performance evaluation of the MINUET system to assess the effectiveness and efficiency in supporting the monitoring and dissemination of urban emergency events over the time. Our evaluation was taken by simulation. We implemented MINUET in C++ programming language and the simulation was carried out using NS3, version 3.28, along with SUMO. We used IEEE 812.11p with 5.9GHz frequency band, 10MHz bandwidth and bitrate of 6Mbps in the simulation. Furthermore, two snippets of the LuST (Luxembourg SUMO Traffic) were used as traffic mobility scenarios, where traffic patterns follow the actual data of the city of Luxembourg. DCA \cite{tal_novel_2016} and PCTT \cite{khakpour_using_2017} algorithms were chosen to perform the vehicular clustering, although other techniques can also be used. They employ an one-hop and a multi-hop approach, respectively. It is worth noting that PCTT creates groups only with vehicles that detect the event. Simulations were performed in four scenarios, considering a fixed event, different vehicle speeds, road direction and distance between the EB and the event. These scenarios allows us to assess how such variations influence MINUET performance. In all scenarios there is only one event that lasts for 10 minutes.

\begin{figure}[!h]
	\centering
	\includegraphics[scale=0.42]{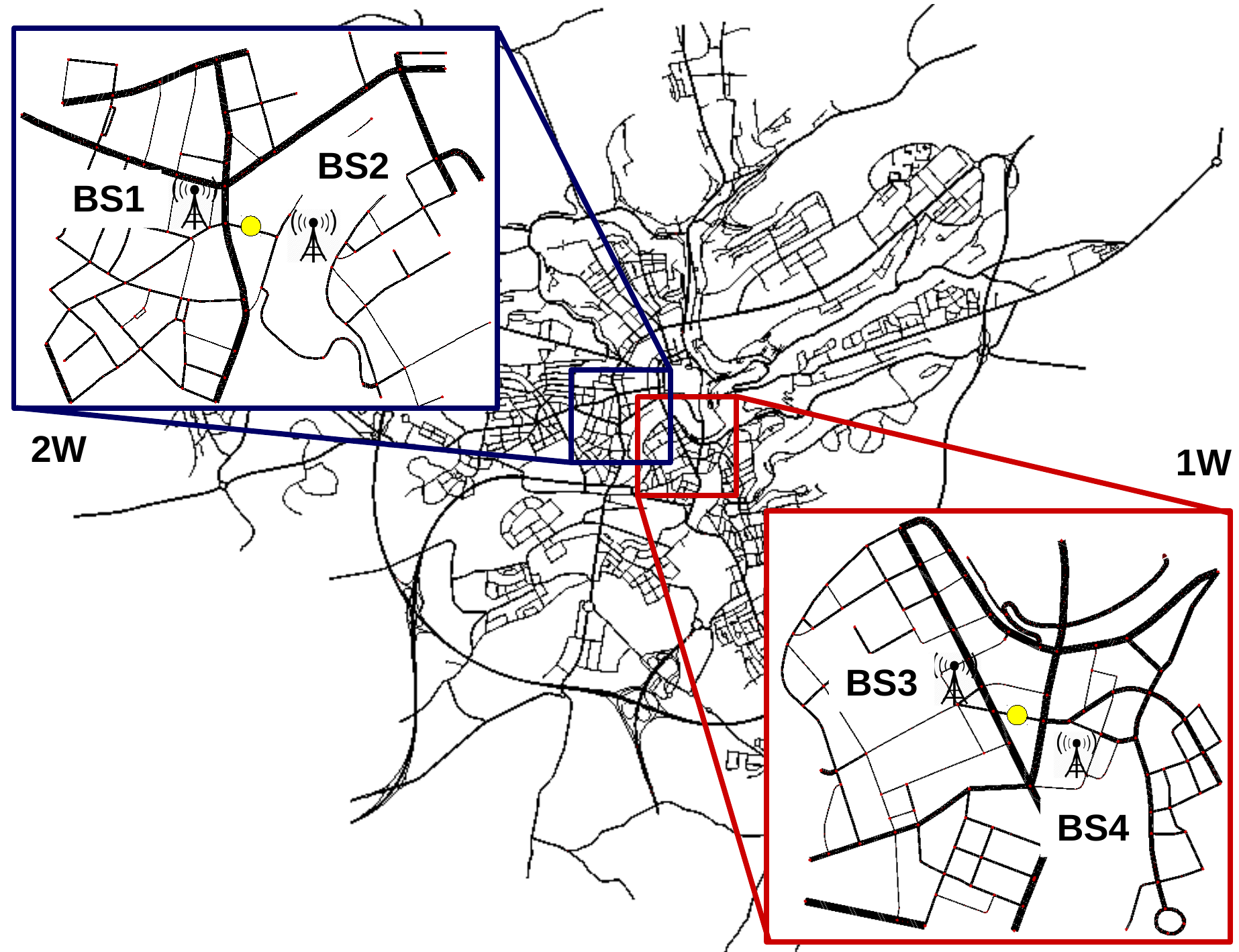}
	\caption{LuST simulation regions}
	\label{fig:scenarios}
\vspace*{-5mm}
\end{figure}

Our simulations take into account LuST regions of 1km$^{2}$ as shown in Figure \ref{fig:scenarios}. We have simulated two scenarios in each region, dealing with distinct time slots to reach different vehicle densities. The yellow dots point out the position where the critical events arise. In the blue region the event shows up in a two-way (2W) road, while in the red region, the event appears in an one-way (1W) road. The snippets also exhibit the positions of the BSs in each region. Vehicles taking part in the monitoring group can travel both on the road where the event is located and on nearby roads. Table \ref{tab:params} summarizes the setting parameters used in the simulations of the four scenarios, which are: (i) 2WLD (two-way, low density); (ii) 2WHD (two-way, high density); (iii) 1WLD (one-way, low density) and; (iv) 1WHD (one-way, high density).
\label{sec:eval}

\begin{table}[h]
\renewcommand{\arraystretch}{0.8}
\centering
\footnotesize
\caption{Simulation settings}
\label{tab:params}
\begin{tabular}{@{\extracolsep{2pt}}lllll}
\hline
\textbf{Parameter} & \textbf{2WLD} & \textbf{2WHD} & \textbf{1WLD} & \textbf{1WHD} \\ \hline 
Sim. time slot (h:m) & 6:28-6:38 & 8:20-8:30 & 6:28-6:38 & 8:20-8:30 \\
Number 
of lanes & 2 & 2 & 2& 2 \\
Kind of lane &  two-way & two-way & one-way & one-way\\
Density (Vehicles/Km) & 16.09 & 31.82 & 16.69 & 45.66 \\
Transmission range (m) & 100 & 100 & 100 & 100 \\
\hline
\end{tabular}
\vspace{-4mm}
\end{table}

To assess the effectiveness of the system, we applied metrics that measure the efficacy (\textit{N\textordmasculine of Detecting Vehicles}, \textit{N\textordmasculine of Generated Monitoring Packets} and \textit{N\textordmasculine of Delivered Monitoring Packets}) and efficiency (\textit{Network Message Overhead}, \textit{Delivery Rate}, \textit{Redundancy Rate} and \textit{Average Delivery Delay}) of the MINUET and the performance of clustering techniques (\textit{Number of Clusters}, \textit{Clustered Vehicles Rate} and \textit{Clustering Overhead}), as follows:
\vspace{-2mm}

\begin{itemize}
\item \textbf{N\textordmasculine of Detecting Vehicles} (DV($\Delta$t)): Indicates the amount of vehicles in the AZ of a given event that detected and monitored such event in each measured time interval, being achieved by Equation \ref{eq:dv}, where $N$ is the total number of vehicles within the AZ in a $\Delta t$ time interval. It points out the collaborative monitoring degree provided by MINUET over time.
\begin{equation}
\footnotesize
    DV(\Delta t) = \sum_{i=1}^{N} 1, \forall \textnormal{ vehicle } v_i \textnormal{ where } detect(v_i) = true
\label{eq:dv}
\end{equation}

    \item \textbf{N\textordmasculine of Generated Monitoring Packets} (MP\textsubscript{gen}($\Delta$t)): Means the total amount of packets yielded for monitoring of a given event ($P_{gen}$) by every detecting vehicle ($v_i$) in each measured time interval. This metric assists to asses the MINUET feasibility in monitoring such event.
\begin{equation}
\footnotesize
    MP_{gen}(\Delta t) = \sum_{i=1}^{DV(\Delta t)} (P_{gen}(v_i))
\label{eq:adr}
\end{equation}

    \item \textbf{ N\textordmasculine of Delivered Monitoring Packets} (MP\textsubscript{deliv}($\Delta$t)): Means the total amount of monitoring packets delivered to a given BS in each measured time interval. It is calculated by Equation \ref{eq:deliv}, where N is the total packets being transmitted in a $\Delta t$ time interval. Along with the former metric, they enable to analyse the MINUET efficacy in delivering monitoring data to external entities.
\begin{equation}
\footnotesize
    MP_{deliv}(\Delta t) = \sum_{i=0}^{N(\Delta t)} 1, \forall \textnormal{ packet } p_i \textnormal{ delivered to a given BS }
\label{eq:deliv}
\end{equation}

    \item \textbf{Network Message Overhead} (NMO): Corresponds to the communication cost by running the MINUET. It takes into account the monitoring ($MP_{transm}$), announcement ($AP_{transm}$) and clustering ($CP_{transm}$) packets transmitted by all vehicles in a given time interval ($V(\Delta t)$).
\begin{equation}
\footnotesize
    NMO = \sum_{i=1}^{V(\Delta t)} (MP_{transm}(v_i)) + \sum_{i=1}^{V(\Delta t)} (AP_{transm}(v_i)) + \sum_{i=1}^{V(\Delta t)} (CP_{transm}(v_i))
\label{eq:adr}
\end{equation}

    \item \textbf{Delivery Rate} (TxD): Corresponds to the rate of the monitoring packets successfully delivered to the BS, obtained by the total number of delivered packets (\textit{MP\textsubscript{deliv}}) divided by the total number of generated packets (\textit{MP\textsubscript{gen}}).  
\begin{equation}
\footnotesize
    TxD = \frac{MP_{deliv}}{MP_{gen}} * 100
\label{eq:adr}
\end{equation}

    \item \textbf{Redundancy Rate} (TxR): Corresponds to the rate of redundant monitoring packets delivered to the BS. It consists of the total copies of the packets delivered more than once (\textit{MP\textsubscript{Ddeliv}}), due to the broadcast operation, divided by the total number of delivered packets (\textit{MP\textsubscript{deliv}}).  
\begin{equation}
\footnotesize
    TxR = \frac{MP_{Ddeliv}}{MP_{deliv}} * 100
\label{eq:adr}
\end{equation}

    \item \textbf{Average Delivery Delay} (ADD): Denotes to the average delay of the monitoring packets delivered to the BS, being calculated by Equation \ref{eq:add}, where $p_i$ is a packet $i$, $Tdeliv(p_i)$ is the delivery time of packet $i$ and the $Tgen(p_i)$ is the generation time of packet $i$.
\begin{equation}
\footnotesize
    ADD = \frac{\sum_{i=1}^{N} (Tdeliv(p_i) - Tgen(p_i))}{MP_{deliv}}
\label{eq:add}
\end{equation}

    \item \textbf{Number of Clusters} (NC): Indicates the total of clusters (C) created over the monitoring period $[t_0, t_f]$ of a given event.
\begin{equation}
\footnotesize
    NC = \sum_{\forall C_it_m, C_jt_n}\delta_{ij}= 
    \begin{cases} 
    1 \textnormal{ if } i \neq j\\
    0 \textnormal{ if } i = j 
    \end{cases} 
    \textnormal{ where } t_m, t_n \in [t_o, t_f]
\label{eq:cvr}
\end{equation}

    \item \textbf{Clustered Vehicles Rate} (TxCV): Means the percentage of vehicles that participated in any group over the monitoring period of a given event, obtained by the total number of clustered vehicles ($CV_{total}$) divided by the total number of vehicles ($NV_{total}$).
\begin{equation}
\footnotesize
    TxCV = \frac{CV_{total}}{NV_{total}} * 100
\label{eq:cvr}
\end{equation}

    \item \textbf{Clustering Overhead} (CO): Denotes the overhead imposed by the exchange of clustering messages, being represented by the percentage of transmitted clustering packets to the total transmitted packets generated by the MINUET operation.
\begin{equation}
\footnotesize
CO = \frac{CP_{total}}{(MP_{total} + AP_{total} + CP_{total})} * 100
\label{eq:co}
\end{equation}
\end{itemize}

\subsection{Analysis of the MINUET performance}
\begin{figure}[h]
    \centering
    \subcaptionbox{DCA-1WHD\label{fig:ncv-a}}{\includegraphics[width=1.3in]{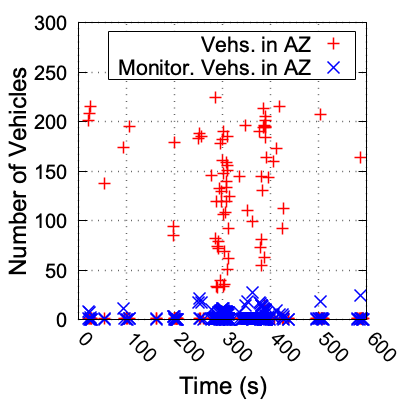}}%
    \subcaptionbox{PCTT-1WHD\label{fig:ncv-b}}{\includegraphics[width=1.3in]{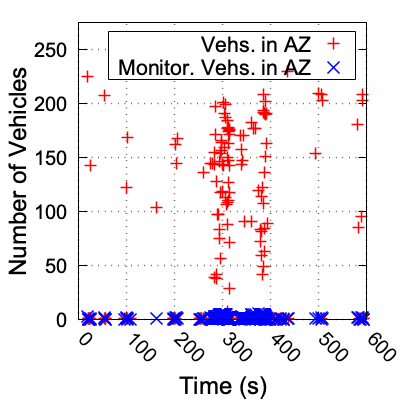}}%
    \subcaptionbox{DCA-1WLD\label{fig:ncv-c}}{\includegraphics[width=1.3in]{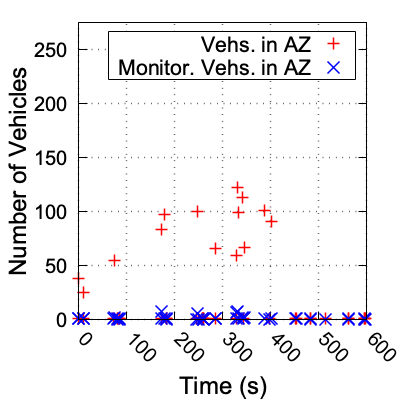}}
    \subcaptionbox{PCTT-1WLD\label{fig:ncv-d}}{\includegraphics[width=1.3in]{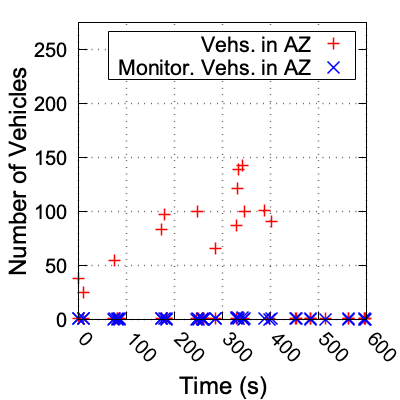}}\\
    \subcaptionbox{DCA-2WHD\label{fig:ncv-e}}{\includegraphics[width=1.3in]{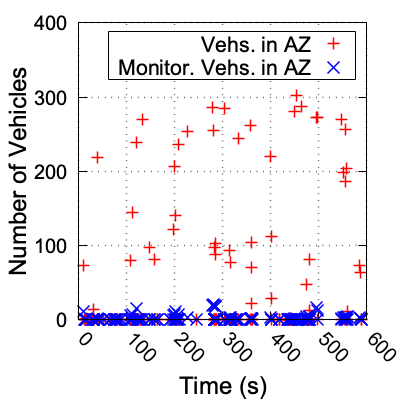}}
    \subcaptionbox{PCTT-2WHD\label{fig:ncv-f}}{\includegraphics[width=1.3in]{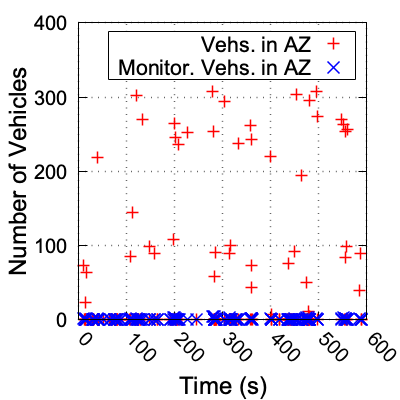}}
    \subcaptionbox{DCA-2WLD\label{fig:ncv-g}}{\includegraphics[width=1.3in]{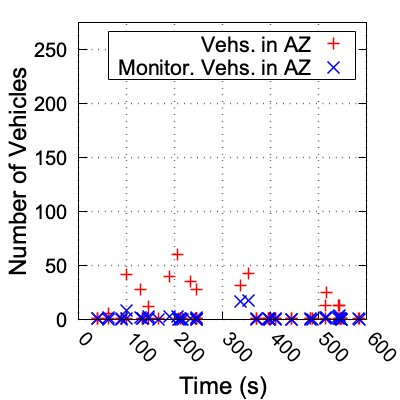}}
    \subcaptionbox{PCTT-2WLD\label{fig:ncv-h}}{\includegraphics[width=1.3in]{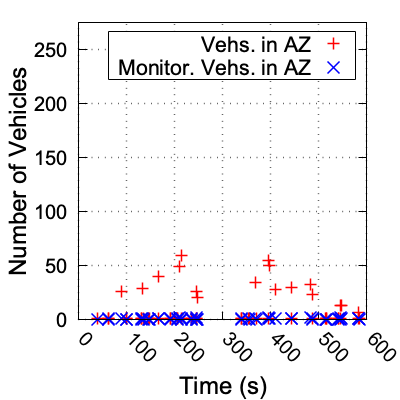}}
\caption{Cooperation of vehicles in the AZ}
\vspace{-5mm}
\label{fig:ncv}
\end{figure}

Figure \ref{fig:ncv} shows that MINUET succeeds in supporting a collaborative monitoring when the event is detected. The graphs of Figure \ref{fig:ncv} present the number of vehicles in the AZ when the event is detected and the number of those vehicles that detect and monitor the event. It is easy to see that there are more than one vehicle simultaneously monitoring the event in several occasions, thus demonstrating the collaborative monitoring capability of MINUET. In addition, there are more vehicles monitoring in higher density scenarios (Figures \ref{fig:ncv-a}, \ref{fig:ncv-b}, \ref{fig:ncv-e} and \ref{fig:ncv-f}) than in the lower density scenarios, as expected. Nevertheless, the results show that MINUET uses the potential of crowdsensing to support the cooperation of vehicles.

\begin{figure}[h]
    \centering
    \subcaptionbox{DCA-1WHD\label{fig:nmpc-a}}{\includegraphics[width=1.3in]{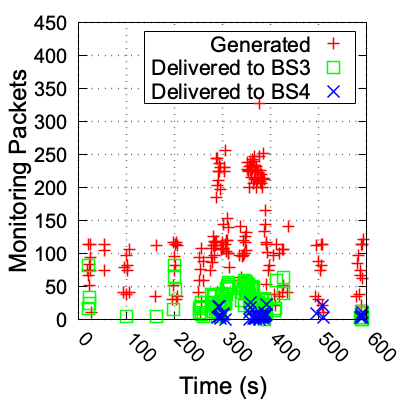}}%
    \subcaptionbox{PCTT-1WHD\label{fig:nmpc-b}}{\includegraphics[width=1.3in]{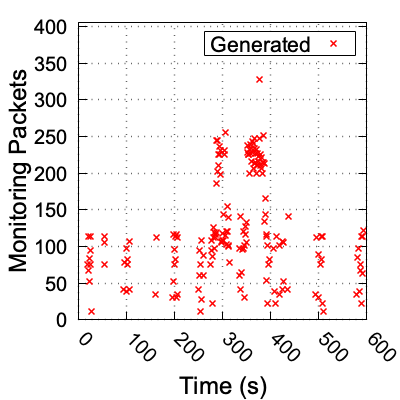}}%
    \subcaptionbox{DCA-1WLD\label{fig:nmpc-c}}{\includegraphics[width=1.3in]{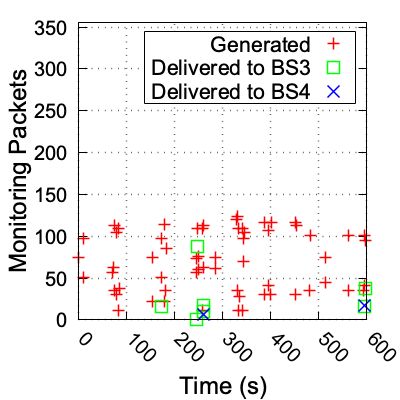}}
    \subcaptionbox{PCTT-1WLD\label{fig:nmpc-d}}{\includegraphics[width=1.3in]{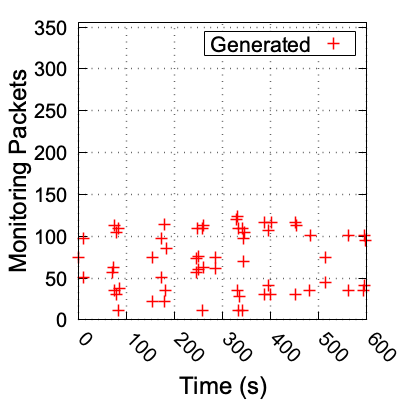}}\\
    \subcaptionbox{DCA-2WHD\label{fig:nmpc-e}}{\includegraphics[width=1.3in]{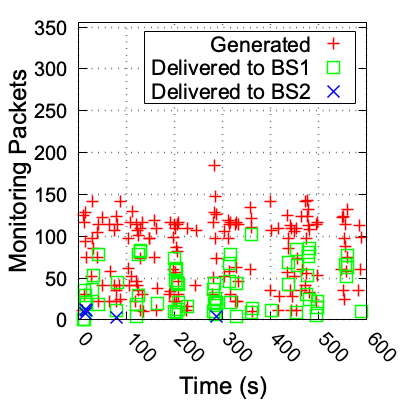}}
    \subcaptionbox{PCTT-2WHD\label{fig:nmpc-f}}{\includegraphics[width=1.3in]{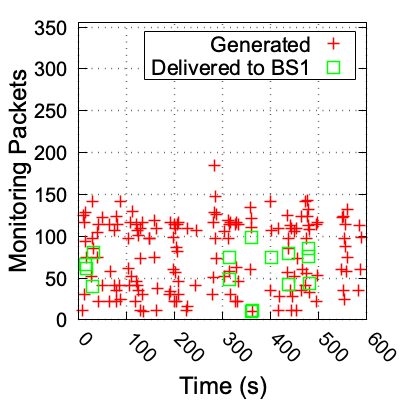}}
    \subcaptionbox{DCA-2WLD\label{fig:nmpc-g}}{\includegraphics[width=1.3in]{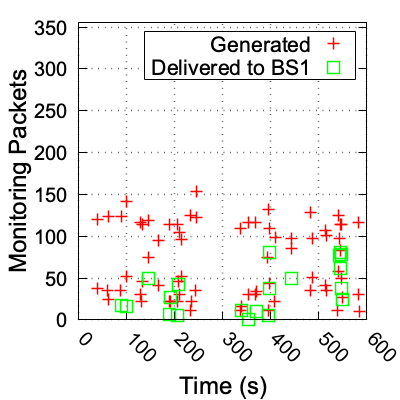}}
    \subcaptionbox{PCTT-2WLD\label{fig:nmpc-h}}{\includegraphics[width=1.3in]{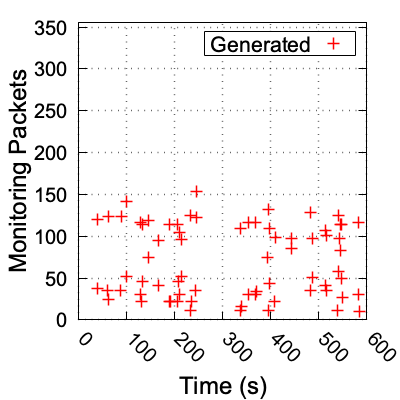}}
\caption{Monitoring packets generated and delivered to the BSs}
\label{fig:nmpc}
\end{figure}

Figure \ref{fig:nmpc} presents the number of monitoring packets generated and delivered to each BS. The results confirm that, once the event is detected, MINUET supports the VANET in delivering the monitoring packets to the BSs. It is worth noting that the numbers of monitoring packets generated are the same when comparing the same scenarios but using different clustering techniques (e.g. Figures \ref{fig:nmpc-a} and \ref{fig:nmpc-b}). This is in accordance to the MINUET behaviour, since the creation of monitoring packets do not depend on the employed clustering technique. When employing DCA, MINUET delivers monitoring packets in all scenarios. On the other hand, when using PCTT, monitoring packets are delivered only in the 2WHD scenario (Figure \ref{fig:nmpc-f}). Those results can be explained by the fact that only vehicles that detected the event at least once can be part of the group in PCTT. Since not all vehicles detect the event, smaller and fewer groups are created. Consequently, less vehicles cooperate and less packets are delivered.

\begin{figure}
    \centering
    \subcaptionbox{DCA-1WHD\label{fig:nmpt-a}}{\includegraphics[width=1.3in]{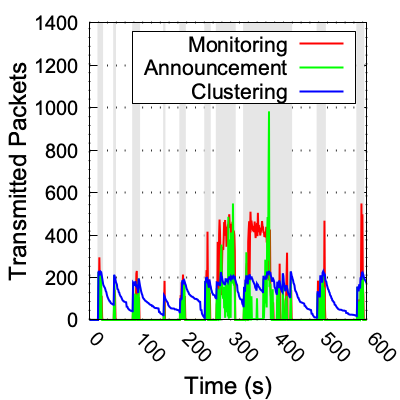}}%
    \subcaptionbox{PCTT-1WHD\label{fig:nmpt-b}}{\includegraphics[width=1.3in]{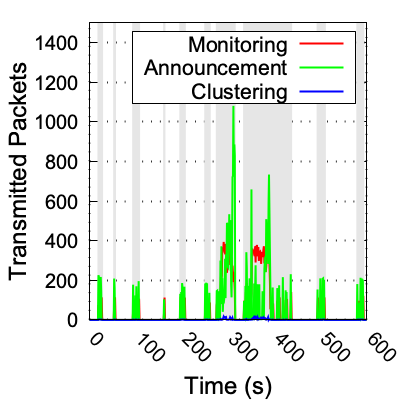}}%
    \subcaptionbox{DCA-1WLD\label{fig:nmpt-c}}{\includegraphics[width=1.3in]{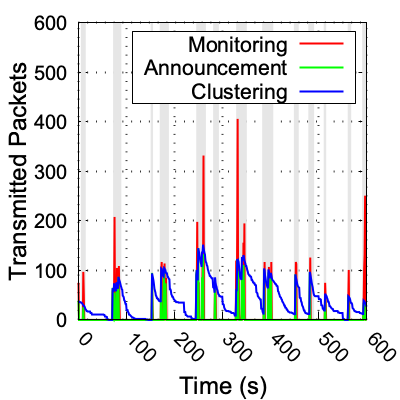}}
    \subcaptionbox{PCTT-1WLD\label{fig:nmpt-d}}{\includegraphics[width=1.3in]{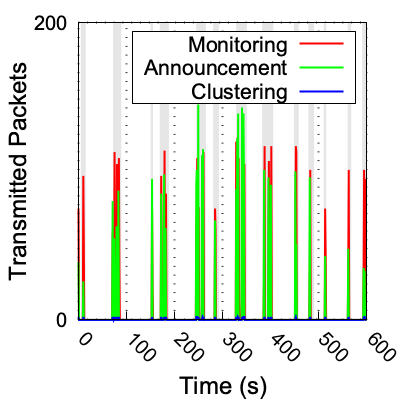}}\\
    \subcaptionbox{DCA-2WHD\label{fig:nmpt-e}}{\includegraphics[width=1.3in]{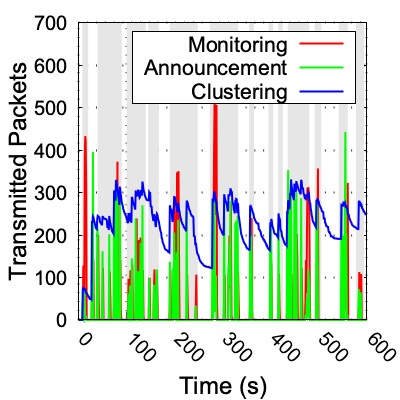}}
    \subcaptionbox{PCTT-2WHD\label{fig:nmpt-f}}{\includegraphics[width=1.3in]{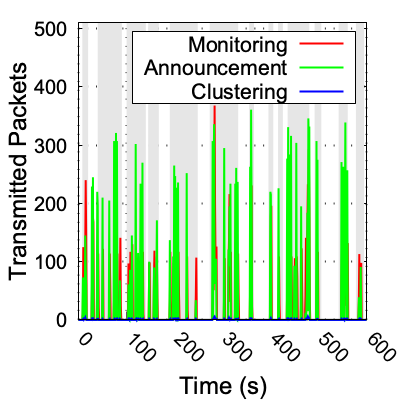}}
    \subcaptionbox{DCA-2WLD\label{fig:nmpt-g}}{\includegraphics[width=1.3in]{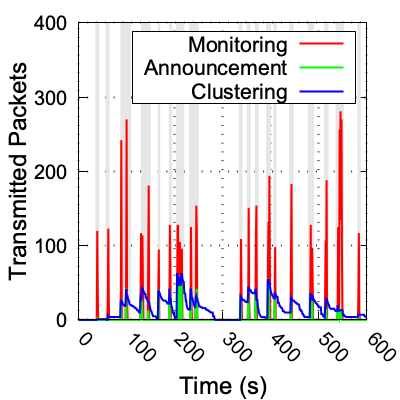}}
    \subcaptionbox{PCTT-2WLD\label{fig:nmpt-h}}{\includegraphics[width=1.3in]{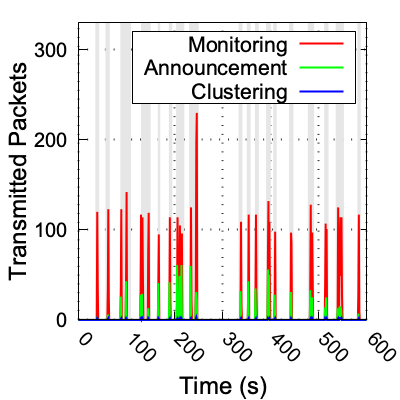}}
\caption{Communication cost by running the MINUET}
\vspace{-5mm}
\label{fig:nmpt}
\end{figure}

Table \ref{tab:rates} fullfills results of Figure \ref{fig:nmpt} by presenting the total number of monitoring packets created and the rates of delivered (\textit{TxD}) and redundant (\textit{TxR}) monitoring packets. By using DCA, the scenarios 2WHD and 1WLD earned the highest and the lowest rates, respectively, as expected. The probability to deliver packets increases if there are more vehicles (i.e. a more connected network). The higher losses of packets arise when groups have one member only. As soon as an event is detected by a vehicle, it starts the cluster formation and the dissemination of the data. Since there are no other members in the group in the early clustering stages, no monitoring packets are received by the BSs. Nonetheless, the \textit{TxR} gained about 50\% in three scenarios, thus increasing the availability of data in the EE.

\begin{table}[]
\renewcommand{\arraystretch}{0.8}
\centering
\footnotesize
\caption{Delivered and redundant monitoring packets}
\label{tab:rates}
\begin{tabular}{|l||c|c|c||c|c|c|}
\hline
& \multicolumn{3}{c||}{\textbf{DCA}} & \multicolumn{3}{c|}{\textbf{PCTT}}\\
\cline{2-4} \cline{5-7}
\textbf{Scenario} & \multicolumn{1}{c|}{\textbf{MP\textsubscript{deliv}}} & \multicolumn{1}{c|}{\textbf{TxD(\%)}} & \multicolumn{1}{c||}{\textbf{TxR(\%)}} & \multicolumn{1}{c|}{\textbf{MP\textsubscript{deliv}}} & \multicolumn{1}{c|}{\textbf{TxD(\%)}} & \multicolumn{1}{c|}{\textbf{TxR(\%)}} \\
\hline
\textbf{1WHD} & 21880 & 17.01 & 55.42 & 21880 & -- & --\\
\textbf{1WLD} & 4281 & 4.95 & 50.11 & 4281 & -- & --\\
\textbf{2WHD} & 12337 & 21.59 & 49.62 & 12337 & 7.21 & 5.11\\
\textbf{2WLD} & 5014 & 14.35 & 35.3 & 5014 & -- & --\\
\hline
\end{tabular}
\vspace{-4mm}
\end{table}

The network communication cost generated by MINUET can be seen in the graphs of Figure~\ref{fig:nmpt}. They show the number of monitoring, announcement and clustering packets that are transmitted in the network. It is important to note that, as the event is detected (represented by the grey areas in the graphs), the number of transmitted packets increases. On the other hand, when the event is not detected, this number is likely to decrease, thus not unnecessarily overloading the network. Concerning the clustering techniques, DCA generates and transmit more packets in the network than PCTT, which is in accordance to the previous results. If we compare such graphs with those from Figure \ref{fig:nmpc} we note that there are more packets being delivered to the BSs when there are more clustering packets being transmitted in the network (which increases the probability to create more groups).  

Table \ref{tab:add} shows the average delivery delay of monitoring packets according to the number of hops. MINUET successfully delivered packets up to 5 hops and up to 3 hops when using DCA and PCTT, respectively. The scenario 2WHD was the only one in which packets were delivered by using PCTT. This is in accordance to the previous results, since few and small groups are created with PCTT. The maximum delay was 111ms, which is a short time to deliver monitoring data of critical events. Such results confirm the MINUET efficiency for real-time monitoring of delay-sensitive data.

Performance results of the clustering techniques are shown in Table \ref{tab:clus-perf}. It presents the NC, TxCV and CO values in each scenario. The DCA technique creates more clusters (NC) than PCTT in all scenarios. The percentage of vehicles that participate in clusters (TxCV) is also higher in DCA than in PCTT. Such results lead to the higher overhead caused by DCA during the clustering process (CO), since there are more clusters being formed with more vehicles. It is worth noting that in the 2WHD scenario, the CO of DCA corresponds to about 80\% of the network communication overhead, which makes the overhead imposed by MINUET (announcement and monitoring messages) reaching only 20\% of the total overhead. Overall, DCA imposes more overhead than PCTT, which is a consequence of the fact that DCA creates more groups in the simulated scenarios, thus disseminating and delivering more packages. Nevertheless, such results attest that MINUET can deal with different clustering techniques to disseminate event information.

\begin{table}[]
\renewcommand{\arraystretch}{0.8}
\caption{Average Delivery Delay (s)}
\label{tab:add}
\footnotesize
\centering
\begin{tabular}{|l||c|c|c|c||c|c|}
\hline
& \multicolumn{4}{c||}{\textbf{DCA}} & \multicolumn{2}{c|}{\textbf{PCTT}} \\
\hline
\textbf{\# 
of hops} & \textbf{2} & \textbf{3} & \textbf{4} & \textbf{5} & \textbf{2} & \textbf{3} \\
\hline
\textbf{1WHD} & 0.071 & 0.081 & 0.085 & 0.076  & --- & ---\\
\hline
\textbf{1WLD} & 0.055 & 0.071 & 0.077 & --- & --- & ---\\
\hline
\textbf{2WHD} & 0.061 & 0.086 & 0.078 & 0.111 & 0.048 & 0.020\\
\hline
\textbf{2WLD} & 0.054 & 0.065 & --- & --- & --- & --- \\
\hline    
\end{tabular}
\end{table}

\begin{table}[]
\renewcommand{\arraystretch}{0.8}
\centering
\footnotesize
\caption{Clustering techniques performance}
\label{tab:clus-perf}
\begin{tabular}{|l||c|c|c||c|c|c|}
\hline
& \multicolumn{3}{c||}{\textbf{DCA}} & \multicolumn{3}{c|}{\textbf{PCTT}}\\
\hline
\textbf{Scenario} & \textbf{NC} & \textbf{TxCV(\%)} & \textbf{CO(\%)} & \textbf{NC} & \textbf{TxCV(\%)} & \textbf{CO(\%)}\\
\hline
\textbf{1WHD} & 3401 & 43.39 & 53.48 & 57 & 0.72 & 1.05\\
\textbf{1WLD} & 1254 & 26.80 & 75.79 & 31 & 0.66 & 1.28\\
\textbf{2WHD} & 2573 & 43.30 & 80.26 & 70 & 1.17 & 0.80\\
\textbf{2WLD} & 514 & 14.19 & 53.71 & 70 & 0.85 & 1.46\\
\hline
\end{tabular}
\vspace{-4mm}
\end{table}

\vspace{-3mm}
\section{Conclusion}
\label{sec:concl}
This article presented the MINUET system for supporting cooperative monitoring of urban emergency events by the means of on-demand vehicles clustering. MINUET enables the accompaniment of random events in time and space in urban environments through the collaborative monitoring and dissemination of the collected data to an external entity. Once in this
entity, the data can be analyzed, producing information to aid the entity in the decision making regarding the resolution of the event. Simulation results have shown that MINUET can detect an event, create groups that collaborate to disseminate monitoring messages, and deliver the messages to a BS through these groups. As future work, we intend to analyze the behavior of MINUET in scenarios with overlapping events in both time and space, considering mobile events, employing other clustering techniques and evaluating with additional performance metrics, such as packet loss, probability reception rate and event perception.
\vspace{-0.2cm}


\bibliography{minuet-arxiv.bib}


\end{document}